%% file: main.tex
\documentclass[journal]{IEEEtran}

\usepackage{cite}
\usepackage{amsmath,amssymb,amsthm,amsfonts}
\usepackage{amsfonts}
\usepackage{amssymb}
\usepackage{algorithmic}
\usepackage{graphicx}
\usepackage{textcomp}
\usepackage{rotating}
\usepackage{makecell}
\usepackage{multirow}
\usepackage{ulem}
\usepackage{color}
\usepackage{ulem} 
\usepackage{subcaption} 
\usepackage{hyperref}
\usepackage{url}
\usepackage{framed}

\usepackage{colortbl}
\definecolor{mygray}{gray}{.9}
\usepackage{multirow,array}
\usepackage{threeparttable}
\usepackage{bigstrut}

\usepackage{comment}

\ifCLASSINFOpdf
\else
\fi
\begin{document}
%
\title{Can Steering Wheel Detect Your Driving Fatigue?}
%
%
%

\author{Jianchao~Lu,
        Xi~Zheng,~\IEEEmembership{Member,~IEEE,}
        Tianyi Zhang,~\IEEEmembership{Member,~IEEE,}
        Michael~Sheng,~\IEEEmembership{Member,~IEEE,}
        Chen Wang,~\IEEEmembership{Member,~IEEE,}
        Jiong~Jin,~\IEEEmembership{Member,~IEEE,}
        Shui~Yu,~\IEEEmembership{Senior Member,~IEEE,}
        Wanlei Zhou,~\IEEEmembership{Senior Member,~IEEE}
\thanks{Jianchao~Lu, Xi~Zheng, Michael~Sheng are with the Department of Computing, Macquarie University, Sydney, Australia (e-mail: jianchao.lu@hdr.mq.edu.au).}
\thanks{Tianyi Zhang is with the School of Computer Science, Harvard University, Cambridge, MA, United States}
\thanks{Chen Wang is with Data61, CSIRO, Sydney, Australia}
\thanks{Jiong~Jin is with the School of Software and Electrical Engineering, Swinburne University of Technology, Melbourne, Australia.}
\thanks{Shui~Yu, Wanlei~Zhou are with the Faculty of Engineering and Information Technology, University of Technology Sydney.}
}
\maketitle

\begin{abstract}
Automated Driving System (ADS) has attracted increasing attention from both industrial and academic communities due to its potential for increasing the safety, mobility and
efficiency of existing transportation systems. The state-of-the-art ADS follows the human-in-the-loop (HITL) design, where the driver’s anomalous behaviour is closely monitored by the system. Though many approaches have been proposed for detecting driver fatigue, they largely depend on vehicle driving parameters and facial features, which lacks reliability. Approaches using physiological based sensors (e.g., electroencephalogram or electrocardiogram) are either too clumsy to wear or impractical to install. In this paper, we propose a novel driver fatigue detection method by embedding surface electromyography (sEMG) sensors on a steering wheel. Compared with the 
existing
methods, our approach is able to collect bio-signals in a non-intrusive way and detect driver fatigue at an earlier stage. 
The 
experimental 
results show that our approach outperforms existing methods with the weighted average F1 scores about 90\%. We also propose promising future directions to deploy this approach in real-life settings, such as applying multimodal learning using several supplementary sensors.
\end{abstract}


%
\IEEEpeerreviewmaketitle

\input{introduction.tex}

\input{related_work}
\input{background}
\input{methodology}

\input{empirical_setting}

\input{experiment}

\input{conclusion}



\bibliographystyle{IEEEtran}
\normalem
\bibliography{reference}








\end{document}

%% file: introduction.tex
\section{Introduction}

The Automated Driving System (ADS) refers to an automated driving mechanism that takes over the vehicle and allows human drivers to leave all responsibilities to the driving system. 
Several companies have been actively implementing Level 3 ADS projects, meaning that the vehicles can guide themselves automatically under certain conditions~\cite{audi-mediacenter.com}.
However, driver-less vehicles are still far away from us. 
Alternatively, 
ADS is currently adopting a human-in-the-loop (HITL) design, where the driver's anomaly needs to be detected~\cite{litman2019autonomous}.

Detecting anomalous driving behavior is not only important for designing automated driving systems, but also is critical for driving safety.
One of the most serious anomalous driving behaviors is {\em fatigue}, which can be referred as a state where the person is neither in sleep nor awake state~\cite{henni2018feature}. Fatigue driving has become one of the major causes for deaths and accidents across the world. According to NHTSA, around $50,000$ injuries and $800$ deaths were reported from car accidents because of the fatigue driving in the United States alone in 2017~\cite{NHTSA}. Similarly, it accounts $25$\%-$30$\% of the total road accidents in China \cite{huang2018driver}. The ever increasing number of accidents caused by fatigue motivates us to propose a practical yet accurate driving fatigue detection system, which could be eventually integrated into the ADS. 

Several approaches have been proposed for detecting driving fatigue, such as the vehicle behaviour based method. In~\cite{sahayadhas2012detecting}, the authors offer fatigue detection systems by monitoring vehicle movements, including steering wheel angle and lane deviation. However, the shortcoming of such systems is that drivers are typically warned only at a deep fatigue state, which could be too late when the warning occurs. The facial-feature based 
method is another way to detect fatigue driving. For example, 
blink frequency is extracted to evaluate the drivers' fatigue state in~\cite{bergasa2006real}. But such methods may fail to detect fatigue driving because of the surrounding context such as sunlight or darkness, and wearable devices. Moreover, previous work also uses bio-signals for fatigue driving detection.
Electromyography (EMG), electroencephalogram (EEG), electrocardiogram (ECG), and electrooculography (EOG) are typical bio-signals to measure the physiological state of a driver~\cite{huang2018driver}. 
It is, however, difficult to collect data from such bio-sensors 
since they require to attach electrodes and wires directly to the driver. This requirement of attaching sensors to human body while driving is not practical for real-world fatigue detection~\cite{zhang2019driver}. Some recent work also explores the possibility of attaching sensors to driving seats for fatigue detection~\cite{wang2017new}. But in such a scenario, a driver needs to wear thin clothing to facilitate direct contact with the sensors, making such technique impractical in real life. To overcome these limitations, our system adopts surface electromyography (sEMG) sensors, which are attached on the steering wheel of a vehicle for convenient and practical fatigue detection.

In this paper, we propose a steering-wheel based sensor deployment solution to detect fatigue driving degree. Compared with existing methods, our approach is able to (1) keep monitoring the drivers fatigue state while driving rather than a later warning only at a deep fatigue state, (2) avoid failing to detect fatigue driving because of the surrounding context, 
and 
(3) collect the bio-signals in a non-intrusive way. The experiment shows that our approach outperforms existing methods with the weighted average F1 scores about 90\%. In addition, we conduct two real-life 
qualitative studies. The studies show that there are insights and promising future directions to make our solution a reality. Our primary contributions are five-fold:

\begin{itemize}
\item[1)] \textbf{Novel concept:} We propose the idea of using steering-wheel based sensor of deploying bio-signals sensor on steering wheel to detect driver's fatigue driving to address the data collection issues with bio-signals.
\item[2)] \textbf{Tailored sEMG sensors:} We design and build our own sensors together with signal collection device.
\item[3)] \textbf{Sound methodology:} We propose a novel approach called VeSEM to extract valid sEMG signals from noisy data and design two-layer features for the underlying sEMG signals.
\item[4)] \textbf{Extensive studies:} We conduct two extensive studies (including a real-world study involving four experienced drivers) and the results show useful insights to adopt our approach for real-life usage. 
\item[5)] \textbf{New insights:} Based on our study, we point out some promising future research directions for using sEMG sensors to detect driving fatigue, which can be used by other ubiquitous computing applications in general.
\end{itemize}

The rest of the paper is organized as follows. Section~\ref{sec:realted work} discusses the related work of fatigue driving detection systems. Section~\ref{sec:background} walks through the unique characteristics of sEMG sensors. The overview of our proposed method is presented in Section \ref{sec:methodology}. The empirical settings are discussed in Section \ref{sec:empirical_setting}. The experimental results are 
reported 
in Section \ref{sec:results}. Finally, the conclusion and future works are drawn in Section~\ref{sec:conclusion}.

%% file: related_work.tex
\section{RELATED WORK} \label{sec:realted work}
\subsection{Vehicle Behavior based Methods}
When getting fatigue, a driver’s ability to perceive the surrounding traffic circumstance and judge the driving situation will decrease. This change impacts how a driver controls the vehicle, which can be reflected by the vehicle’s abnormal performance~\cite{sahayadhas2014electromyogram}. Therefore, vehicle behaviors collected by various types of sensors inside a vehicle (e.g., accelerometer and gyroscope embedded in the steering wheel) can be used to determine whether the driver is in the fatigue driving. The major methods include vehicle speed detection, steering wheel angle detection, brake pedal force detection and the accelerator pedal force detection~\cite{miyaji2009driver}\cite{victor2005sensitivity}\cite{liu2009predicting}\cite{ engstrom2005effects}\cite{charlton2009driving}. Vehicle behavior based methods can easily collect the data from sensors embedded in the car without affecting the driver's normal driving. However, its detection accuracy is easily affected by driver's driving habits, the weather, road and traffic conditions and other external factors. Moreover, the method is only able to detect the fatigue driving when the driver is about to lose the control of vehicle, which is obviously not safe for the driver~\cite{sahayadhas2012detecting}. Therefore, the fatigue detection result from this method is better to be used as supplementary information for the driver.

\subsection{Facial-feature based Methods}
The techniques based on facial features have been extensively adopted in fatigue driving detection, mainly due to their non-disturbance on the driver's attention in the driving process. Most existing methods attempt to detect drivers' fatigue facial features, e.g., yawning~\cite{zhang2015driver}, blink frequency~\cite{bergasa2006real}, gaze direction~\cite{dasgupta2013vision}, eye state and head position~\cite{mandal2016towards}. In most cases, such methods are capable of recognizing fatigue facial features. However, they may easily fail to detect fatigue driving because of the surrounding context such as sunlight or darkness, and wearable devices~\cite{Azim2009Automatic}. Zhang et al.~\cite{zhang2017driver} reported that glasses disturbed the detection of eye state. Moreover, the shape of eyes in the camera changed significantly during the head rotation~\cite{jo2011vision}. Therefore, it becomes difficult to recognize the state of eyes for any fatigue detection. The current facial-feature based fatigue driving methods lack precise gaze-estimation algorithms to detect head orientation aligning with eye movement. Also, they fail to differentiate the closed eyes state caused by fatigue or vigorous laughter. Another major weakness of the existing methods is their attempt to recognize expressions from high resolution facial images that need to be generated from a controlled environment \cite{tian2004evaluation}. However, in the real scenario, the surveillance images are often of low resolutions, making it more difficult to recognize expressions in this setting.

\subsection{Physiological based Methods}
Physiological based methods are 
widely used 
for fatigue driving detection. They are built to effectively evaluate the fatigue symptoms of drivers based on the bio-signals, including electroencephalogram (EEG) \cite{ chen2015automatic}{\cite{chai2016driver}}, electrooculography (EOG) \cite{barua2019automatic} \cite{huo2016driving}, electrocardiogram (ECG) \cite{sahayadhas2012detecting} \cite{huang2018detection}, and surface electromyography (sEMG) \cite{wang2017new} \cite{fu2014detection} collected by real-time portable sensors. Although physiological based methods have exhibited optimum performance in FDD, it is intrusive in nature~\cite{pramanick2018defatigue} and drivers may be simply not willing to drive with many sensors attached to different parts of their bodies.   

Another main issue with physiological based methods is the difficulty of collecting the bio-signals in a non-contact way (usually electrodes and wires need to be in direct contact with drivers). The drivers are inevitably disturbed by the measuring electrodes. In order to address this issue, in \cite{wang2017new}, sEMG and ECG are collected from the sensors that were deployed in a cushion on a driver’s seat. Similarly, 
a European project called \emph{HARKEN} 
developed a sensor system built into a safety belt and seat cover of cars, which is able to detect fatigue driving behaviour \cite{IEEESPECTRUMSeatbelt}. The proposed method of reading signals in \cite{wang2017new} requires a driver to wear cloth of thickness less than 2 mm.
Besides, there are also methods to use sEMG sensors to analyze muscle signals and detect fatigue driving~\cite{bernardo2018muscle}. However, they require sEMG sensors to be directly attached on the human hands, skin or neck, which may not be practical \cite{pramanick2018defatigue}. In contrast to the existing approaches, we apply sEMG sensors on the steering-wheel to avoid attaching senors on different parts of a human body. 

In summary, there are no practical and reliable solutions for the fatigue driving detection. Therefore, we propose the idea of using sEMG sensor on steering-wheel in order to collect physiological features from bio-sensors for FDD. Our work aims to provide a practical and stable FDD compared to the existing methods. In particular, we use objective quantification of fatigue degree, mainly on the basis of features extracted from the sEMG signals.


%% file: background.tex
\section{Background and Problem Statement}
\label{sec:background}

Most of the sEMG-based fatigue driving detection systems are intrusive,
and 
only a few studies are based on non-intrusive detection systems but they usually have specific requirements, such as the thickness of clothes (less than 2 mm) in~\cite{wang2017new}.  In order to detect fatigue driving using non-intrusive and sEMG based detection systems with minimal special requirements, a new method 
needs 
to be proposed. Considering the exposed skin on palms and fingertips are the most frequent contact part of human body with the steering wheel, detecting fatigue driving through the sEMG sensors attached to a steering wheel could be a new feasible way. 
Similar design could be found in gym and health clubs, where sensors are attached to the handles of fitness equipment. For example, pulse heart rate sensors are attached to the handgrip of treadmills. The limitations of this kind of 
design are that: (1) the valid sensor reading requires users to keep their hands very still. However, it is not an easy thing to do if a user is running especially at peak intensity; (2) handlebar fixed with sensors is awkward for running posture and is discouraged. People could not run naturally if they have their arms placed in front of them and not swinging arms naturally by their side. 

For the steering wheel based sEMG sensor, though it is a lot more intuitive and easier to use and deploy, we have greater problems comparing with the existing deployment (e.g., handlebar in fitness equipment): 
\begin{enumerate}
  \item {\bf Problem 1 (P1):} When drivers change a lane, or make a turning, it is difficult to ask them to hold the steering wheel still or constant contact with the steering wheel. 
  Such resultant hand movement causes sEMG signals distortion or even loss.
  \item {\bf  Problem 2 (P2):} Since different drivers may have different driving habits, in order to make the driving behaviour naturally, we cannot fix the position of a sEMG sensor on a steering wheel.
\end{enumerate}

These problems derive some particular research challenges of fatigue driving detection as follows:
\begin{enumerate}
  \item {\bf Challenge 1 (C1):} sEMG signals distortion or loss caused by {\bf P1} increases the difficulty of valid sEMG signals acquisition from the raw signal. On the one hand, distorted signals or lost signals (represented as a flat reading) are noise signals which must be removed from the raw signals. On the other hand, as the position of a sEMG sensor is not fixed on a steering wheel  ({\bf P2}), different drivers may have different holding postures, resulting in various signal patterns of the valid sEMG signals. Therefore, how to differentiate between the valid sEMG signals and the noise signals, especially the distorted signals, is one of the key challenges in this study.
  
  \item {\bf Challenge 2 (C2):} In order to obtain distinctive fatigue states from the sEMG signals, the interval of the detection points for such fatigue state (Fatigue Detecting Point: FDP) cannot differ greatly. For example, if we set the first FDP at 15:00 and the second FDP at 15:05, it is clearly not appropriate if we set the third FDP at 15:06, but 15:10 will be reasonable. However, sEMG signals distortion or loss caused by {\bf P1} may disturb the balance of these intervals. Due to the complex road conditions and unpredictable hand movements, the interval between the occurrence of 
  a valid sEMG signal in t$_n$ and t$_{n+1}$ can be significantly different with the one in t$_{n+1}$ and t$_{n+2}$ (``t'' is a specific point of time for detection). Therefore, how to choose the distinctive FDP to make the intervals more reasonable is another key challenge in this study.   
\end{enumerate}

%% file: methodology.tex
\section{Methodology}
\label{sec:methodology}
Our study consists of three main components: (1) signal processing and noise filtering, (2) dynamic fatigue detection point selection, and (3) feature generation. In the first component, we pass the raw sEMG signals to filter the noise using band-pass filters. Even so, there still exist some noises after that (e.g., mechanical noises caused by big movement of the steering wheel). We thus segment the filtered signals into $5$-minute sliding windows and further divide each sliding window into much smaller sub-sliding windows with $50\%$ overlapping. They are then fed into ``valid sEMG sample selection machine'' to identify valid sEMG samples. Afterwards, fatigue detection points are selected using our dynamic fatigue detection point selection method. 
Finally, 
the valid sEMG samples in the fatigue detection points become the input of feature generation component, where $28$ features will be generated based on our two-layer feature generation method.  

\subsection{The Design of sEMG Sensor and Data Collection Device}
\begin{figure*}
    \centering
    \begin{subfigure}[b]{1\textwidth}
    \centering
    \includegraphics[width=0.7\linewidth]{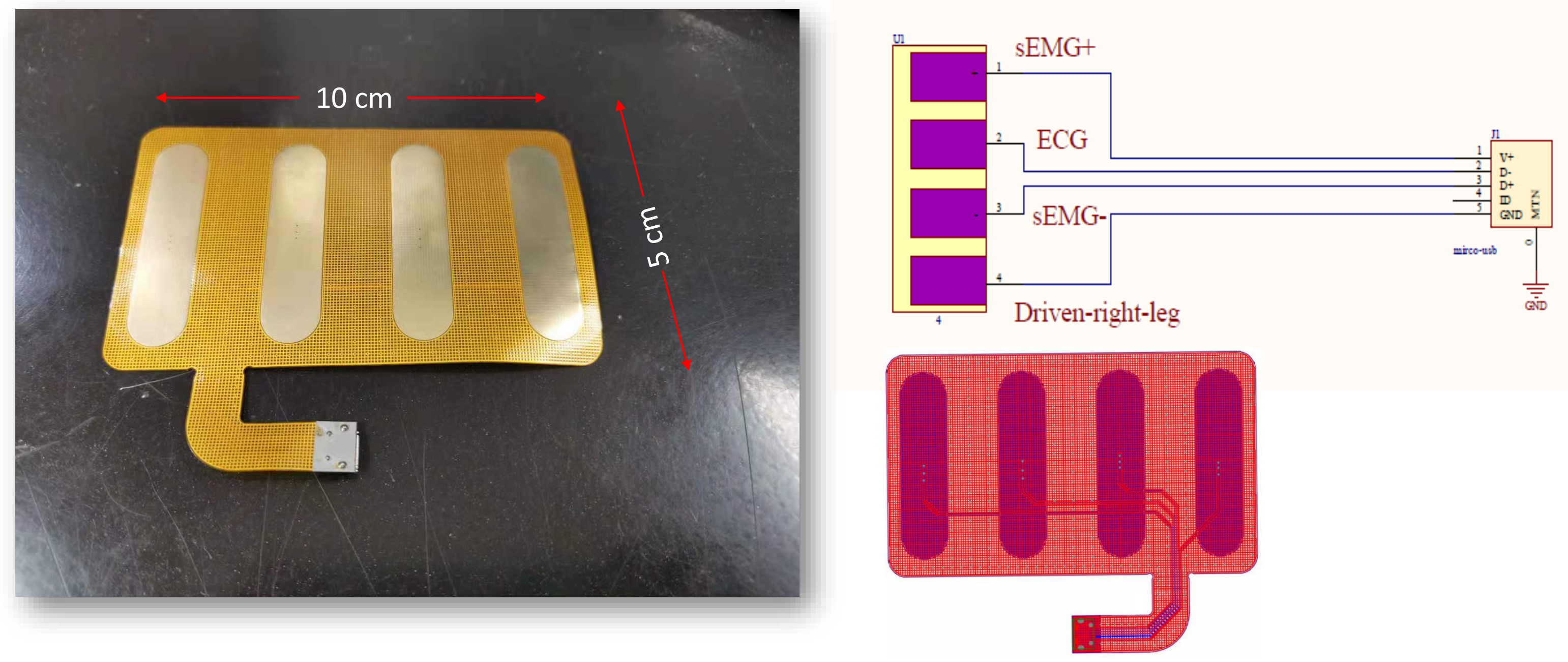}
    \label{sensor}
    \end{subfigure}
    \begin{subfigure}[b]{1\textwidth}
    \centering
    \includegraphics[width=0.7\linewidth]{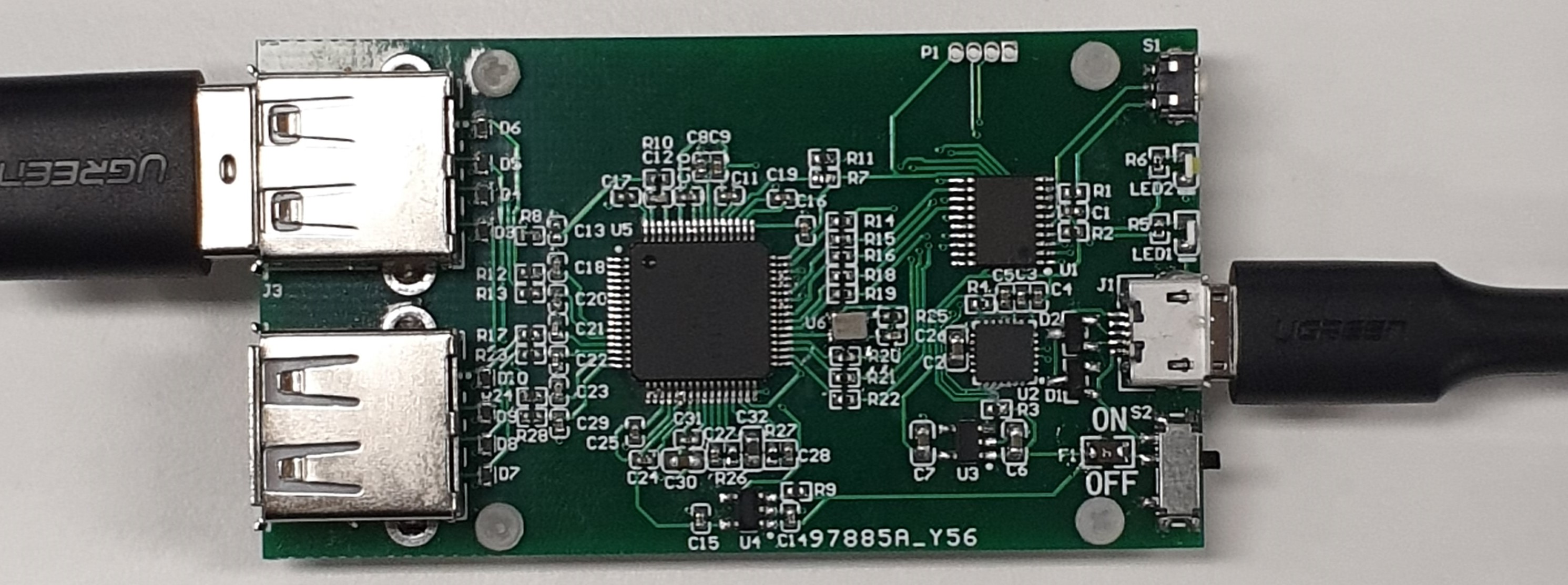}
    \label{board}
    \end{subfigure}
    \caption{The sEMG sensor and acquisition board}
    \label{fig:board}
\end{figure*}

In order to address the 
problem 
{\bf P2} in Section~\ref{sec:background} and satisfy different driving habits, the sEMG sensor is designed with detachable style by using a flexible printed circuit (FPC) board (10 cm $\times$ 5 cm), which easily fits to the shape of the steering wheel. Therefore, drivers can adjust the position of the sensor based on their own habit as shown in Figure~\ref{fig:board}. The signal acquisition electrodes on the sEMG sensor include four strips of copper in each FPC. Two strips are used as positive and negative electrodes to collect the signals from sEMG in a non-contact way. Such a design allows to get the sEMG signal using only one FPC (i.e., one hand). Another strip is ECG electrode, which is designed for our future research of leveraging sEMG with ECG signals for fatigue detection. All the strips of copper are vertical arranged in each FPC so that the palms and fingertips are able to maximum contact with the signal acquisition electrodes.

The data collection board (right side of Figure~\ref{fig:board}) contains two mini USB ports and a micro USB port. The former are connected to two sEMG sensors for left and right hands, and the latter is connected to the computer for transmitting the electric signals. The data collection board is able to collect the sEMG sensor data in real time. 

As compared to the state-of-the-art in detecting driving fatigue, our solution would not disturb the driver’s attention, while traditional approaches either require sEMG sensors to be attached on human body~\cite{pramanick2018defatigue} or to be placed on the cushions of the seats with a thickness requirement on clothing~\cite{wang2017new}. Therefore, our tailored sensor and device are convenient, practical and suitable for detecting driving fatigue in real scenarios.



\subsection{Signal Processing and Filtering}
sEMG signals have low signal-to-noise ratio and are sensitive to noises. In order to remove these noises, we firstly use a fourth order band-pass filter to de-noise the sEMG signals with lower cutoff frequency of 10 $Hz$ and higher cutoff frequency of 300 $Hz$. These two parameters are designed based on the frequency range of sEMG (10 $Hz$-300 $Hz$) \cite{nazmi2016review}\cite{fu2014detection}\cite{von2011removal}\cite{wang2017new}.

\begin{figure}[t]
\centering
\includegraphics[width=3.5 in]{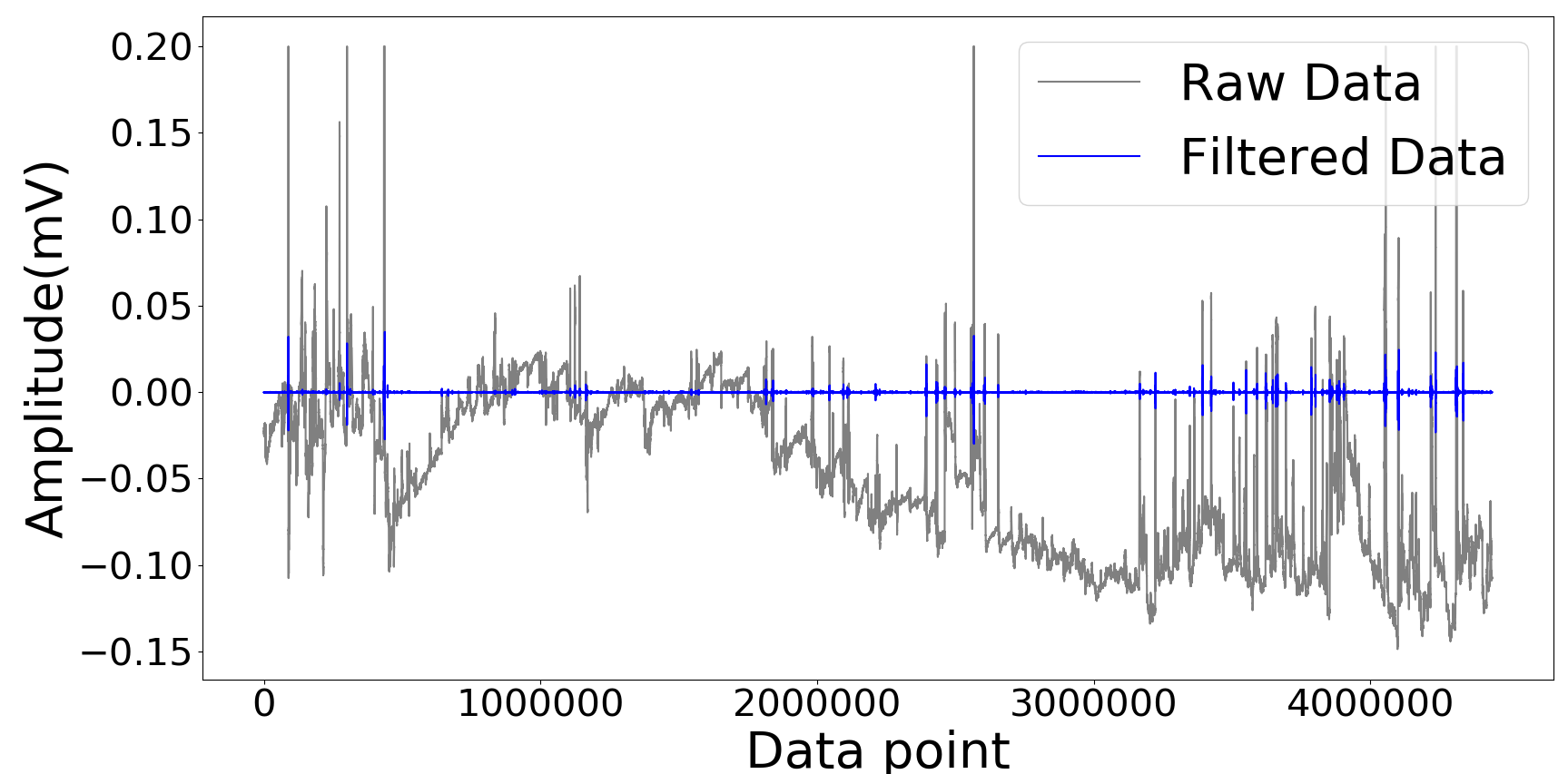}
\caption{Pattern of the sEMG signals before and after filtering }
\label{filtering}
\end{figure}

Figure~\ref{filtering} presents the pattern of sEMG signals before and after filtering. Clearly, the signals are still noisy. These noises are generated due to the 
challenge 
{\bf C1}. In order to further remove noises and obtain valid sEMG signals, we 
design a ``Valid sEMG Selection Machine (VsESM)'', which is a semi-supervised learning based on PU Learning algorithm~\cite{ zhang2018anomaly}. 
\begin{figure*}[t]
\centering
\includegraphics[width=6 in]{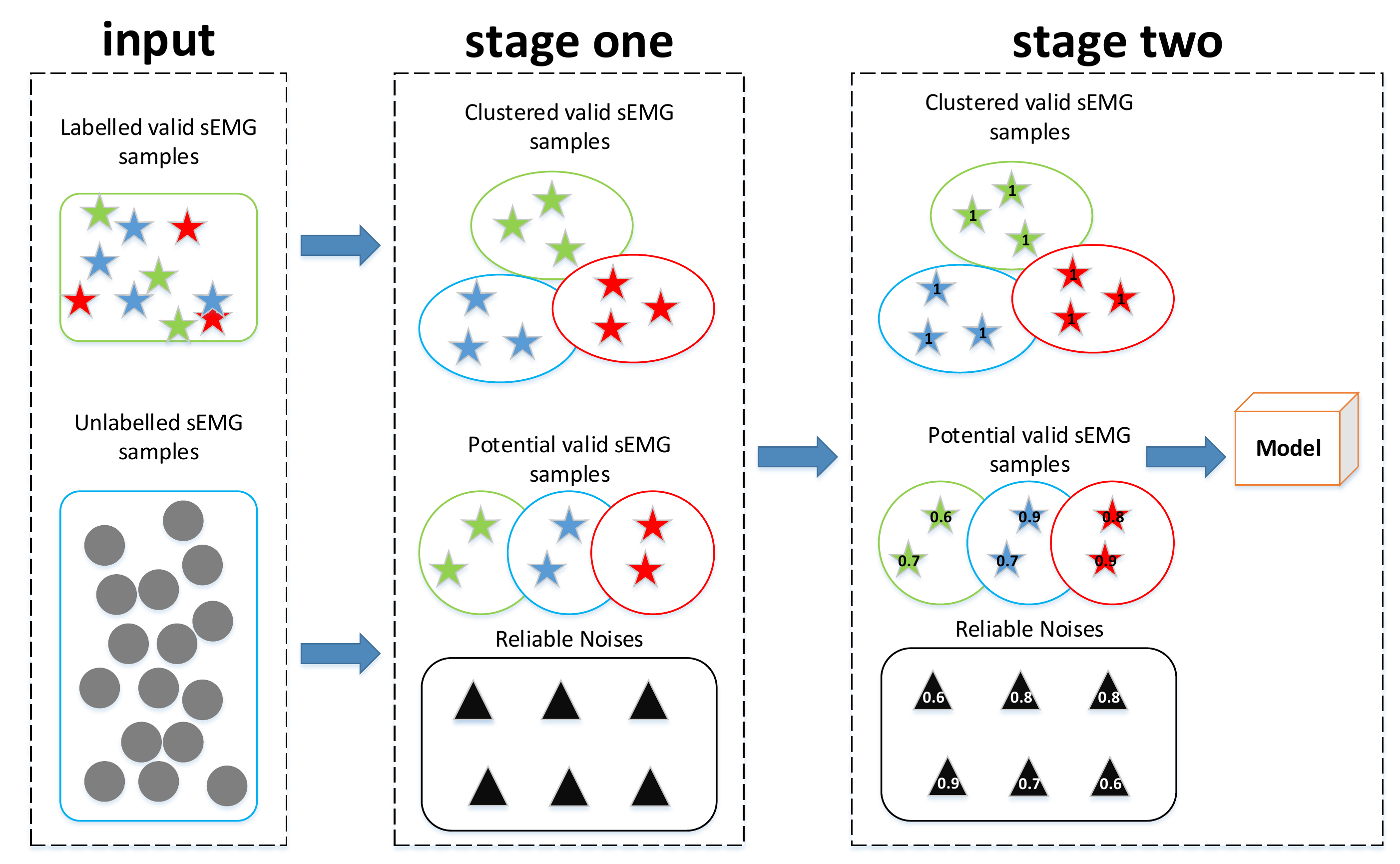}
\caption{The overall framework of VsESM }
\label{VsESM}
\end{figure*}

In the first stage, we prepare two sample sets: labelled valid sEMG sample set and unlabelled sample set. All samples are 
collected based on a two-layer sliding window technique. Generally speaking, driving fatigue is gradually developed as an accumulation process over a period of time, and there is hence no significant change in a relatively short period. However, if the sliding window is set too large, the fatigue detection system may fail to respond timely. In this study, we record the drivers' driving conditions every five minutes and thus we set a 5-minute interval as the window size.  5-minute window size may be too long to extract the clean feature under certain scenarios, Due to significant amount of noises inside sEMG signals deployed on a steering wheel, e.g., mechanical noise, signal shift, electricity interference \cite{fu2014detection} and no sEMG signals at all when drivers' hands are off the steering wheel (e.g., 
a sharp turn of the steering wheel
or turning in a round-about). Therefore, a dual-layer sliding window with a much smaller sub-window with $50\%$ overlapping is adopted in our approach. The total amount of points in one sample is sub-window size $*$ sample frequency data. For example, for a $10$ second sub-window, the amount of points in this sub-window is: $10$ second sub-window size $*$ $1000 Hz$ = 10,000 points. 

And then, both labelled valid sEMG samples and unlabeled samples are used to select valid sEMG samples. We believe that the labelled valid sEMG samples are different to each other, and they should not be simply classified into one cluster. The reason is that the different drivers may have different holding postures, 
making that 
the valid sEMG signals are really diversified. Therefore, we first try to separate them into different clusters, so that the samples in each cluster are similar to each other. For unlabelled samples, we try to filter both potential valid sEMG samples and noise samples from them according to the isolation score and their similarity score to the labelled valid sEMG samples. The intuition is that, on one hand, the potential valid sEMG signals should be different to noise samples (i.e., can be easily isolated); on the other hand, they should be similar to some labelled valid sEMG signals. In the second stage, we build a weighted multi-class model (in this study, we use ``XGBoost''\footnote{https://xgboost.ai/} as the weighted multi-class model) to distinguish different valid sEMG samples from the noise samples. For the labelled valid sEMG samples, the weights are set to 1, and for the filtered samples, the weights are set according to the confidence of their attached labels (i.e., label ``0'': noise, lebel ``1'': valid sEMG samples).

\subsection{Dynamic Fatigue Detection Point Selection}
In order to address the challenge 
{\bf C2}, we propose a dynamic fatigue detection point selection method to make the detection intervals reasonable. Since each 5-minute window frame may contains many valid sEMG samples, and the indices of 
these samples in current 5-minute window frame can be significantly different with the ones in its neighbour frame, we use the shortest absolute distance of index of the valid sEMG samples between two consecutive 5-minute window frames to select the fatigue detection point: 
\begin{equation}
\label{detection_point_E}
\textsf{$(detection~point)_{n+1}$} = arg \min\limits_{index_{n+1}} |\textsf{index}_{n+1}^{(i)} - \textsf{index}_n^{(j)}|\\
\end{equation}
 where $(detection~point)_{n+1}$ is the detection point of the (n+1)$th$ 5-minute window frame, $index_{n+1}^{(i)}$ and $index_{n}^{(j)}$ are the i$th$ index of the valid sEMG samples in the (n+1)$th$ 5-minute window frame and the j$th$ index of the valid sEMG samples in the (n)$th$ 5-minute window frame, respectively. Here, when calculate (n+1)$th$ fatigue detection point, $index_{n}^{(j)}$ is a fixed point which represents the last fatigue detection point.

\begin{figure}[htp]
	\centering
	\includegraphics[width=0.5\textwidth]{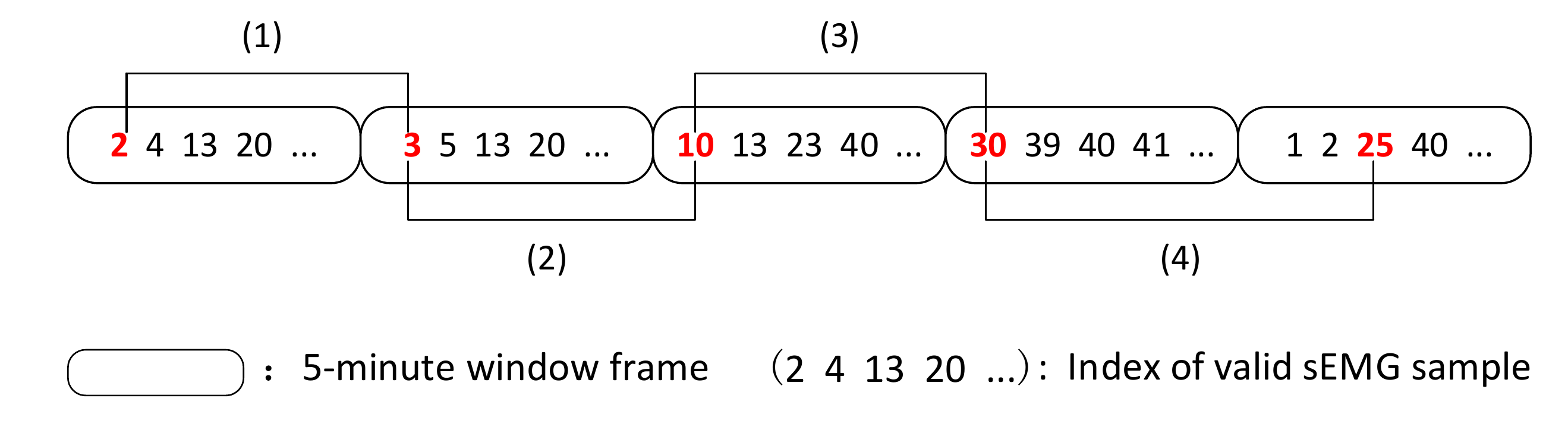}
	\caption{Detection point selection}
	\label{detection-point}
\end{figure}

An example 
is
shown in Figure~\ref{detection-point}. There are five consecutive 5-minute window frames in the figure, where each of frame includes several valid sEMG samples (for example, for the first 5-minute window frame, the data located in the indies of sub-window 2, 4, 13, 20 are the valid sEMG samples). Now, if we set the index "$2$" in the first 5-min window frame as the first fatigue detection point, we can find 
index "$3$" in the second 5-minute window frame has the shortest absolute distance with the index "$2$" in the first 5-min window frame using the Equation~\ref{detection_point_E}. Therefore, we set index "$3$" 
as 
the fatigue detection point for the second 5-minute window frame, by doing so, the most representative feature could be obtained from the valid sEMG sample of index "$3$" in this 5-minutes window. We can use the same way to get the rest detection points, i.e., index "10" in the third 5-min window frames, index "30" in the fourth 5-min window frames, index "20" in the fifth  5-min window frames, respectively. 

\subsection{Feature Generation}
Initially, $14$ features are generated from the detection point obtained from the previous section, including the 12 most commonly used features in previous studies: mean, standard deviation (std), median, maximum (max), minimum (min), the difference between maximum and minimum (max\_min), the signal magnitude area (sma), skewness, kurtosis, zero crossing (time\_over\_zero), mean frequency, median frequency; and two particular features: sample entropy and Lempel-Ziv complexity, which have been used by our most related study~\cite{wang2017new}.

In \cite{wang2017new}, a smooth decreasing tendency on the change of fatigue state is presented using the cushion solution. In order to evaluate whether our solution has a similar tendency with theirs, we draw the tendency chart for all the $14$ features and find all of them present different tendency (ups and downs trend) compared with the physiological features tendency illustrated in~\cite{wang2017new}. 
One explanation for this difference is that when using steering wheel based fatigue driving detection method, drivers tend to make small adjustments of hand holding posture when they feel fatigue. It is intuitive as a body's natural reaction to fatigue. Since a human body movement increases the amplitude of the sEMG signals \cite{mustard1987relationship}, for this natural fatigue recovery behavior, we can find the corresponding unique sEMG signal patterns, a wave trend (down and up). However, the frequency of adjustments on body sitting posture is much less than the frequency of posture adjustments on hand holding posture, and therefore the feature tendency in~\cite{wang2017new} is much smooth than ours. As a result, we cannot directly use the features introduced in ~\cite{wang2017new}.

\begin{figure}[t]
\centering
\includegraphics[width=3.25 in]{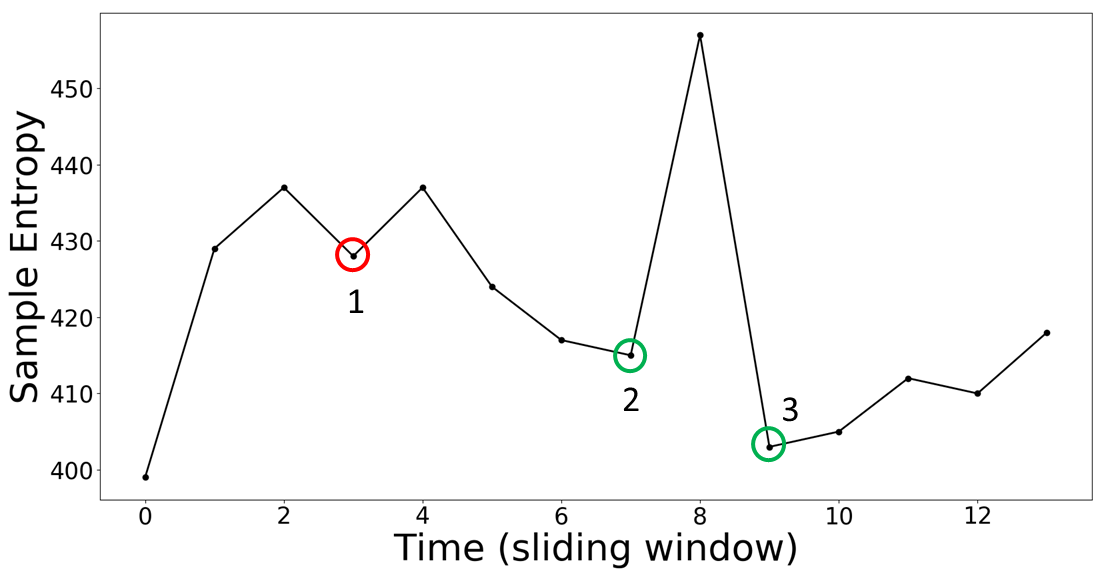}
\caption{Examples of {\em FST} signal pattern in one sEMG feature - Sample Entropy}
\label{up trend}
\end{figure}

In our case, we define a term \textit{fatigue state transition} ({\em FST}), that is, transiting from not feeling fatigue to start feeling fatigue or feeling more fatigue. FST helps us to analyze the drivers' fatigue behaviours. In order to capture those characteristics of sEMG signals corresponding to {\em FST}, we design the second layer sub-features from these wave trends. We first calculate the slope of every two adjacent data points for a given feature to obtain such ``wave'' like patterns. However, in some real-life cases, the rise part of the ``wave" form (the increase of sEMG feature value) could be caused by changing lane, when a driver needs to turn the steering wheel, or some unconscious little finger movements (e.g., some drivers like to rub steering wheel using fingers unintentionally). It is shown as the red circle of point 1 in Figure~\ref{up trend}. To differentiate those from actual FST-related ``wave'' form (points 2 and 3), we use the absolute distance between every two adjacent data points, because the bigger the distance, the more likely the FST-related ``wave'' form is. That is, slope and absolute distance, on top of those four features:
\begin{equation}
\begin{array}{l}
{\textsf{slope}} = ({\textsf{value}_{n+1}} - {\textsf{value}_n}) / ({\textsf{time}_{n+1} - \textsf{time}_n})\\
\textsf{absolute~distance} = |\textsf{value}_{n+1} - \textsf{value}_n|\\
  \end{array}  
\end{equation}
Finally, 
28 features are generated in our study (i.e., $14$ initial features $*$ $2$ slop and absolute distance).

%% file: empirical_setting.tex
\section{Empirical Setting}
\label{sec:empirical_setting}
In order to evaluate the effectiveness and practicality of our sensors and proposed methodology, we conducted two extensive experiments ({\bf E1} and {\bf E2}) in different settings. Both experiments were done with the sEMG sensors installed on a driving steering wheel and the signals were collected with our data acquisition board. The sensor data were stored in a laptop for driver fatigue detection. 

\begin{figure}
	\centering
\includegraphics[width=0.45\textwidth]{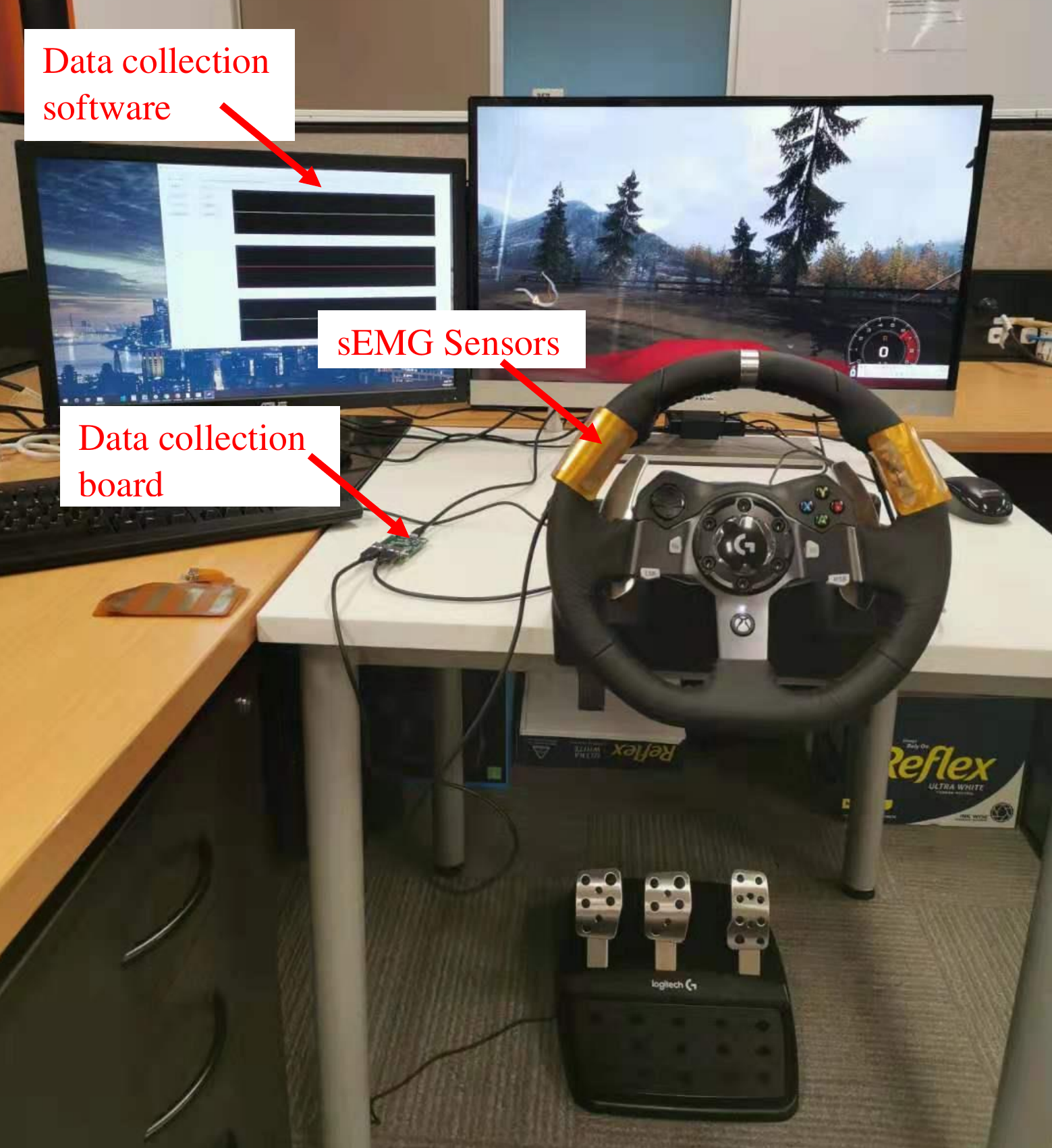}
	\caption{Experimental setting for {\bf E1}}
	\label{lab-test}
\end{figure}

{\bf E1} is conducted in a lab environment, as shown in Figure~\ref{lab-test}. The ambient temperature was set to $26$$^{\circ}$C. We recruited $13$ experienced drivers ($7$ males and $6$ females) with an average of $5$ years' driving experience. All participants were healthy adults of $20$-$30$ years of age, who did not feel any fatigue when they started driving. (1) For training sEMG sample set collection, each participant continuously drove for $1.5$ hours. The experiments were done in four different time slots from $9$ am to $8.30$ pm. More specifically, three participants were asked to perform the experiments from $9$ am to $10.30$ am, three participants from $1$ pm to $2.30$ pm, another three from $4$ pm to $5.30$ pm, and finally four participants from $7$ pm to $8.30$ pm. All the participants were asked to drive freely without any extra constraints on speed and steering movement. (2) For sEMG validation sample set collection, each participant's driving was done in a more controlled way than in the collection of training sEMG sample set. They were asked to hold the steering wheel tightly and maintained a speed between $60$-$80$ km/hr. The controlled driving involved fewer sudden movements of the steering wheel. In addition, participants had a tighter contact with the sensors while driving. All the other settings are identical to the training sEMG sample set collection.

We used the car racing game ``Need for Speed: Hot Pursuit-free driving model''~\cite{nfs11} as the simulation software, and collected the signal data with a sampling frequency of $1000$ Hz. In the free driving model, the participants drove freely in a big city map, even though they need to pay more attention to the driving due to the cars they drove are the high-performing sport cars. This setting makes the participants much easier to get fatigue.

\begin{figure}
    \centering
    \begin{subfigure}{0.5\textwidth}
    \centering
	\includegraphics[width=0.7\linewidth]{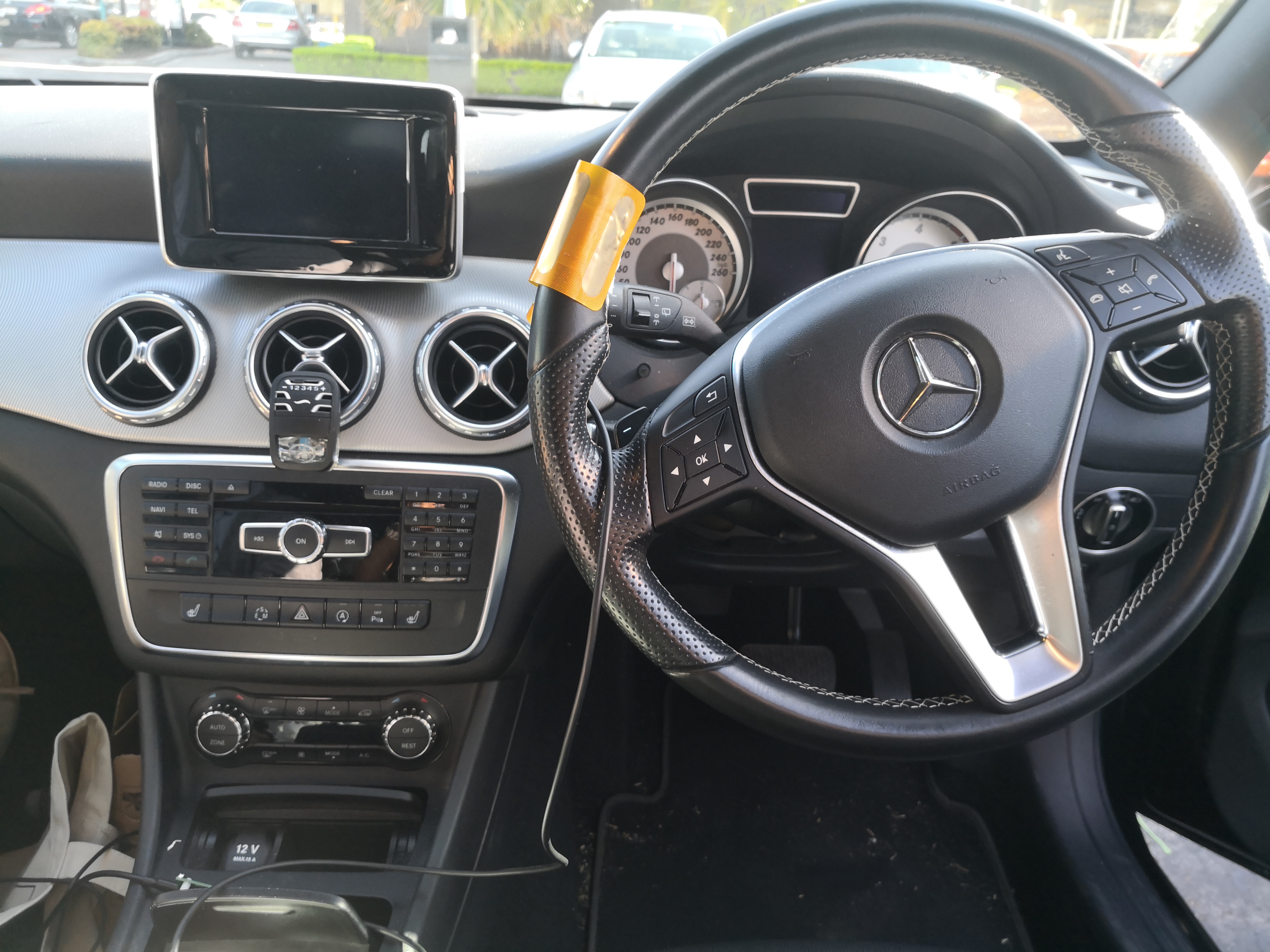} 
	\caption{The driving vehicle}
	\label{real-test}
    \end{subfigure}
    \begin{subfigure}{0.5\textwidth}
    \centering
	\includegraphics[width=0.7\linewidth]{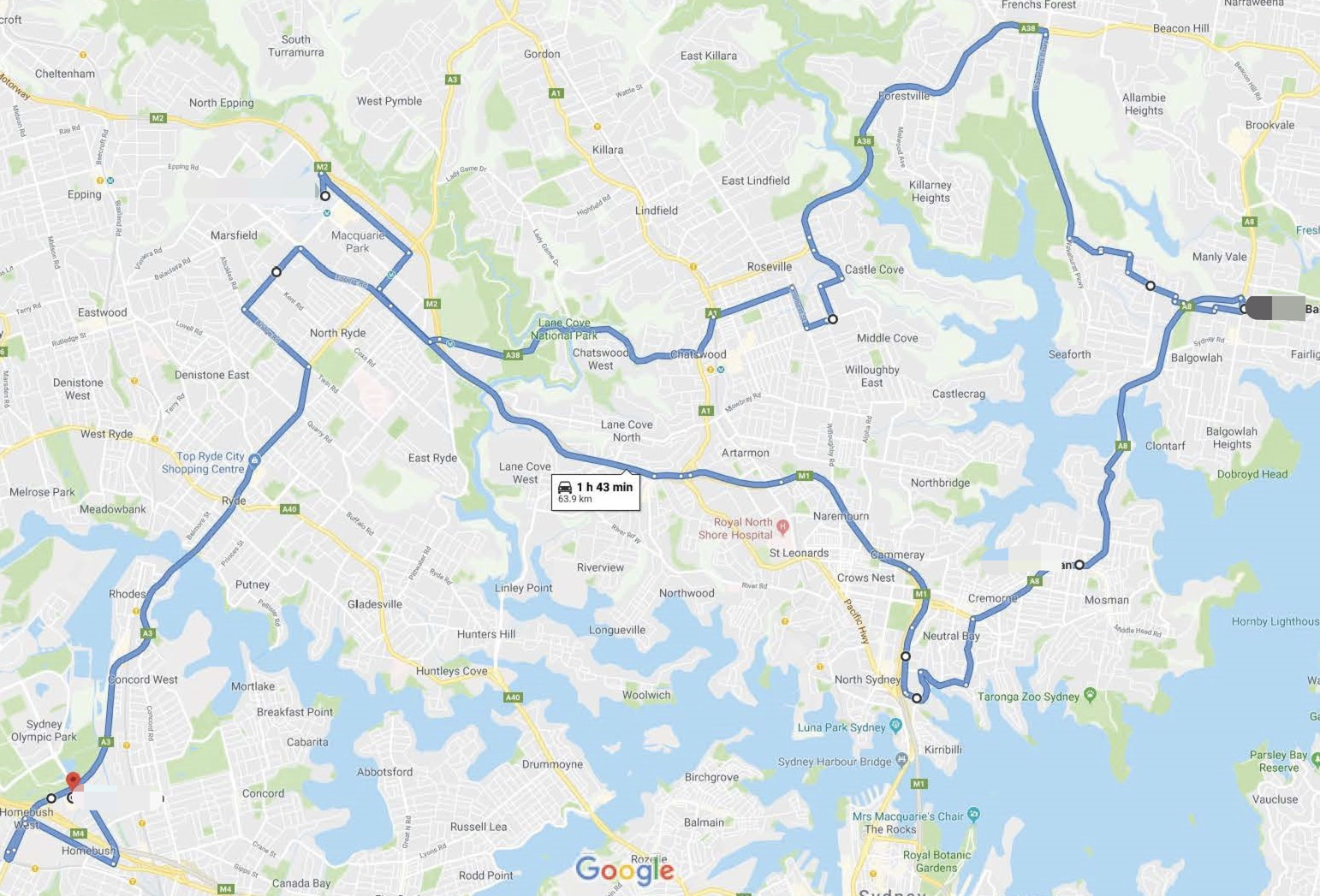}
	\caption{Road-map for driver 1 and driver 2}
	\label{map1}
    \end{subfigure}
    \begin{subfigure}{0.5\textwidth}
    \centering
	\includegraphics[width=0.7\linewidth]{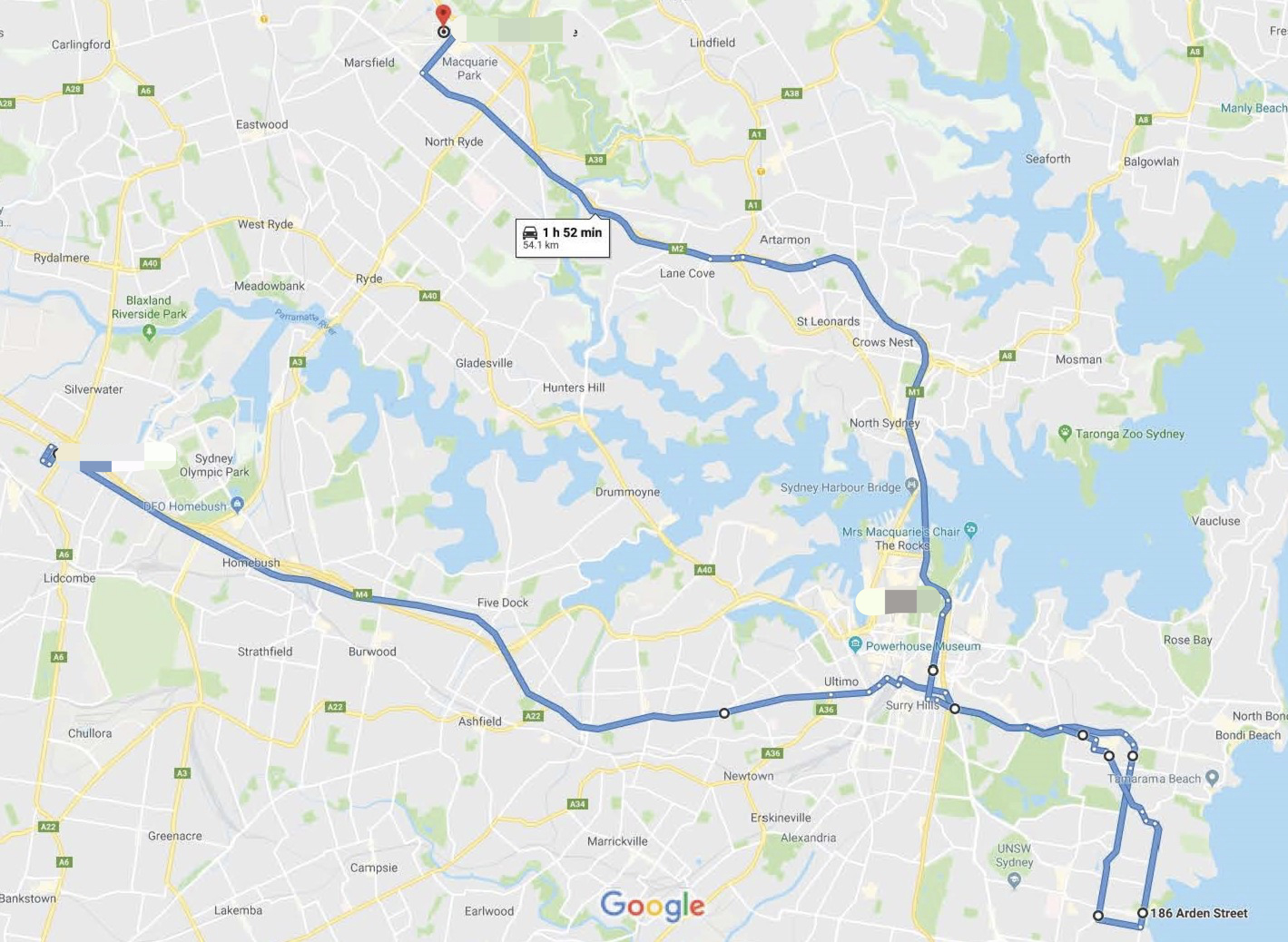}
	\caption{Road-map for driver 3 and driver 4}
	\label{map2}
    \end{subfigure}
    \caption{Experimental setting and the studies for {\bf E2}}
    \label{fig:my_label_}
\end{figure}

For~{\bf E2}, as displayed in Figure~\ref{real-test}, we conducted four on-road tests in a major city in Australia. The cars used were a Mercedez GLA200 2016 Model and 
an
Audi Q7 2011 model with one sEMG sensor attached to the left-hand side of their steering wheel, respectively. The reason for not deploying the sensor to the right-hand side is to avoid dangling the cable connecting our board with the sensor, which may generate more noises. Four male drivers with an average of $10$ years driving experience were asked to drive in four road tests each. All participants were healthy adults of 30-40 years of age. In all real-world tests, the drivers drove freely and reported their fatigue level every $5$ minutes. 
The first two on-road tests were conducted in the morning from $9$ am to $11$ am, as shown in Figure~\ref{map1}.
A total of $63.9$ km driving was done in about $100$ minutes. The second two were conducted at night from $6$ pm to $8$ pm, 
as shown in Figure~\ref{map2}. A total of $54.1$ km driving was done in about $95$ minutes.

In this study, the results of subjective questionnaire ``Swedish Occupational Fatigue Inventory'' (SOFI)~\cite{wang2017new} were used as labels for defining the fatigue state. SOFI intends to measure work-related fatigue by adopting a multidimensional perspective questions, where each question will be ranked from $0$ to $10$. We asked these questions to participants every $5$ minutes and calculated their scores (the higher the score they obtain, the more fatigue the participants they feel). Moreover, we recorded participants facial features using 
the GoPro Hero7 4K Action Cam~\cite{goPro}, which features high video stabilisation, and used a computer vision based fatigue detection system~\cite{zhang2019driver} to get the participants' fatigue state as a benchmark to improve the labels reliability.

%% file: experiment.tex
\section{RESULTS AND DISCUSSION}
\label{sec:results}
\subsection{E1 Result}

\subsubsection{Accuracy Analysis with Fixed Sub-window Size}

In our experiment {\bf E1}, we applied the following $5$ different machine learning algorithms: Support Vector Machine (SVM), Random Forest (RF), Logistic Regression (LR), K-nearest neighbor (KNN) and Naive Bayes (NB). The performance comparison of these models is given in Table~\ref{F1}, where a $30$ second sub-window size is used for feature generation, as the same configurations with the study~\cite{wang2017new}, and the hyper-parameters of models were tuned using the grid search approach. Since there are more NFST (non fatigue state transition) samples than FST (fatigue state transition) ones in our dataset, ``weighted average F1 score'' is used to evaluate our fatigue detection models. Table~\ref{F1} indicates that our random forest based detection model is capable of distinguishing between NFST and FST , with the weighted average F1 score of $96\%$. The confusion matrix for {\bf E1} are given in Table~\ref{Confusion Matrix}, in which $32$ FSTs are correctly identified and $6$ FST is not. The reason behind false positives and false negatives is mainly due to noises during the experiment. The participants in {\bf E1} were asked to drive freely without any speed and steering movement limitations. With no speed limitation, participants tended to drive fast. In the simulation game, it was very difficult to control the vehicle in high speed, so 
a large number of 
big wheel adjustments were performed to control the vehicle. Such significant movements of steering wheel produced 
significant amount of noises in data, thereby causing misclassifications. Also, with no steering wheel movement limitations, participants did not hold the steering wheel very tight. The hands often moved away from the deployed sEMG sensors, producing high motion artefacts that are part of the transient baseline change caused by the electrode motions due to a subject's movement. These motion artefacts created more noises for the underlying sEMG sensor data, thus leading to some false negatives. The experimental results of experiment {\bf E1} showed promising potential of identifying fatigue from our proposed solution. We intended to check further if the accuracy can be affected by the different sub-window size.

\begin{table*}[htp]
\caption{F1 score of various methods with a sub-window size of 30s}
\label{F1}
\centering
\resizebox{\linewidth}{!}{%
\begin{tabular}{|m{1cm}|m{2cm}|m{15cm}|}
\hline
\textbf{Model} &\textbf{F1 score}  &\textbf{Parameters} \bigstrut\\ \hline
SVM&0.71&C=100, cache\_size=200, class\_weight=None, coef0=0.0,
  decision\_function\_shape='ovr', degree=3, gamma='auto\_deprecated',
  kernel='linear', max\_iter=-1, probability=False, random\_state=None,
  shrinking=True, tol=0.001, verbose=False\bigstrut\\ \hline
RF&\textbf{0.96}&bootstrap=True, clas\_weight=None, criterion='entropy',
            max\_depth=25, max\_features='auto', max\_leaf\_nodes=None,
            min\_impurity\_decrease=0.0, min\_impurity\_split=None,
            min\_samples\_leaf=1, min\_samples\_split=2,
            min\_weight\_fraction\_leaf=0.0, n\_estimators=5, n\_jobs=None,
            oob\_score=False, random\_state=None, verbose=0,
            warm\_start=False\bigstrut\\ \hline
LR&0.70&C=100, class\_weight=None, dual=False, fi\_intercept=True,
          intercept\_scaling=1, max\_iter=100, multi\_class='warn',
          n\_jobs=None, penalty='l2', random\_state=None, solver='warn',
          tol=0.0001, verbose=0, warm\_start=False\bigstrut\\ \hline
KNN&0.74&algorithm='auto', leaf\_size=30, metric='minkowski',
           metric\_params=None, n\_jobs=None, n\_neighbors=9, p=2,
           weights='uniform'\bigstrut\\ \hline
NB&0.73&alpha=1.0, binarize=0.0, class\_prior=None, fit\_prior=True\bigstrut\\ \hline
\end{tabular}
}
\end{table*} 

\begin{table}[htp]
\caption{RF Confusion Matrix (E1)}
\label{Confusion Matrix}
\centering
\resizebox{1\linewidth}{!}{%
\begin{tabular}{|m{2cm}|>{\centering\arraybackslash}p{3cm}|>{\centering\arraybackslash}p{2cm}|}
\hline
&Detected NFST& Detected FST\bigstrut\\ \hline
Actual NFST&141&1\bigstrut\\ \hline
Actual FST&6&32\bigstrut\\ \hline
\end{tabular}
}
\end{table} 

\subsubsection{Accuracy Analysis with Various Sub-window Size}
\begin{figure}[t]
\centering
\includegraphics[width=3.3 in]{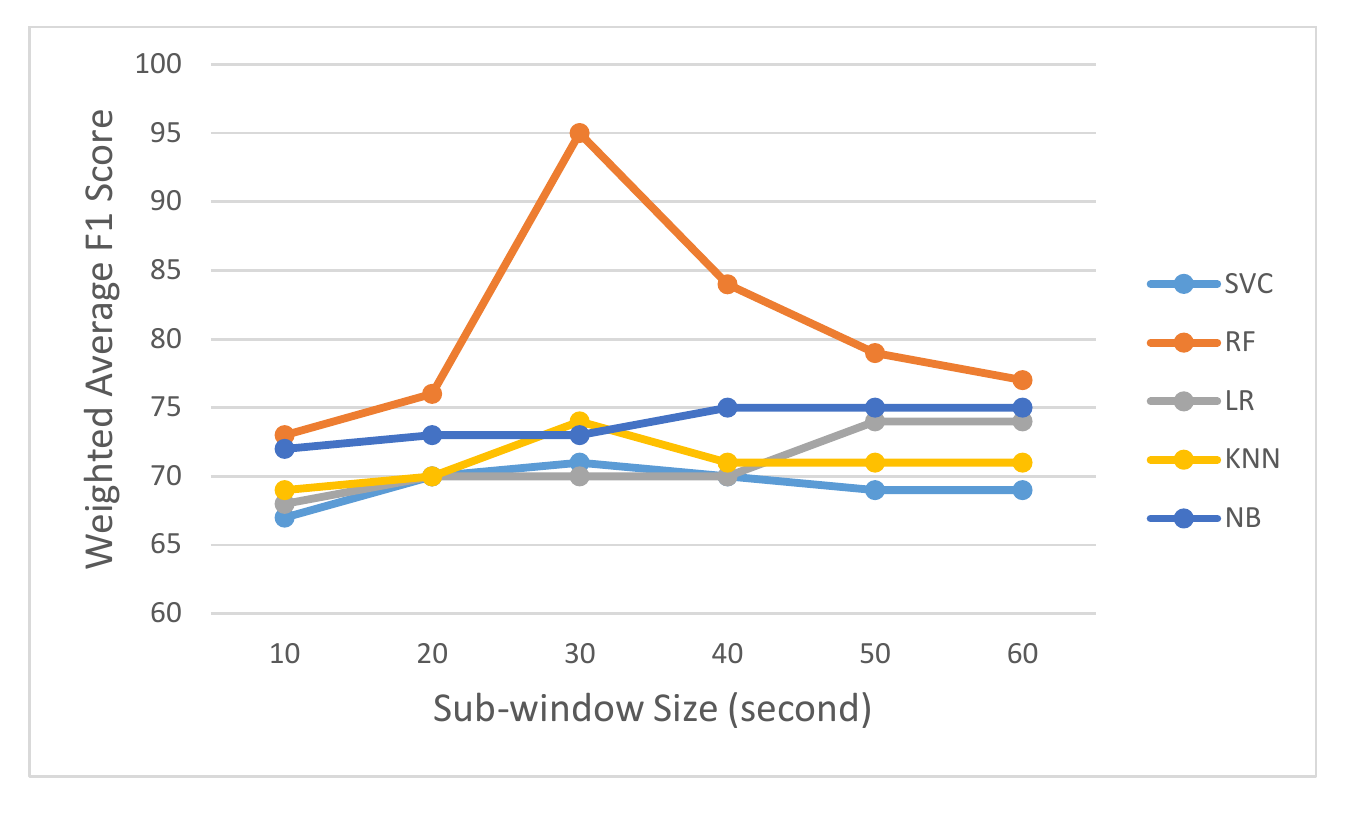}
\caption{Effect of the sub-window size on the fatigue driving detection model performance (F1-score)}
\label{sub_window_effect}
\end{figure}

\begin{figure}[t]
\centering
\includegraphics[width=3.3 in]{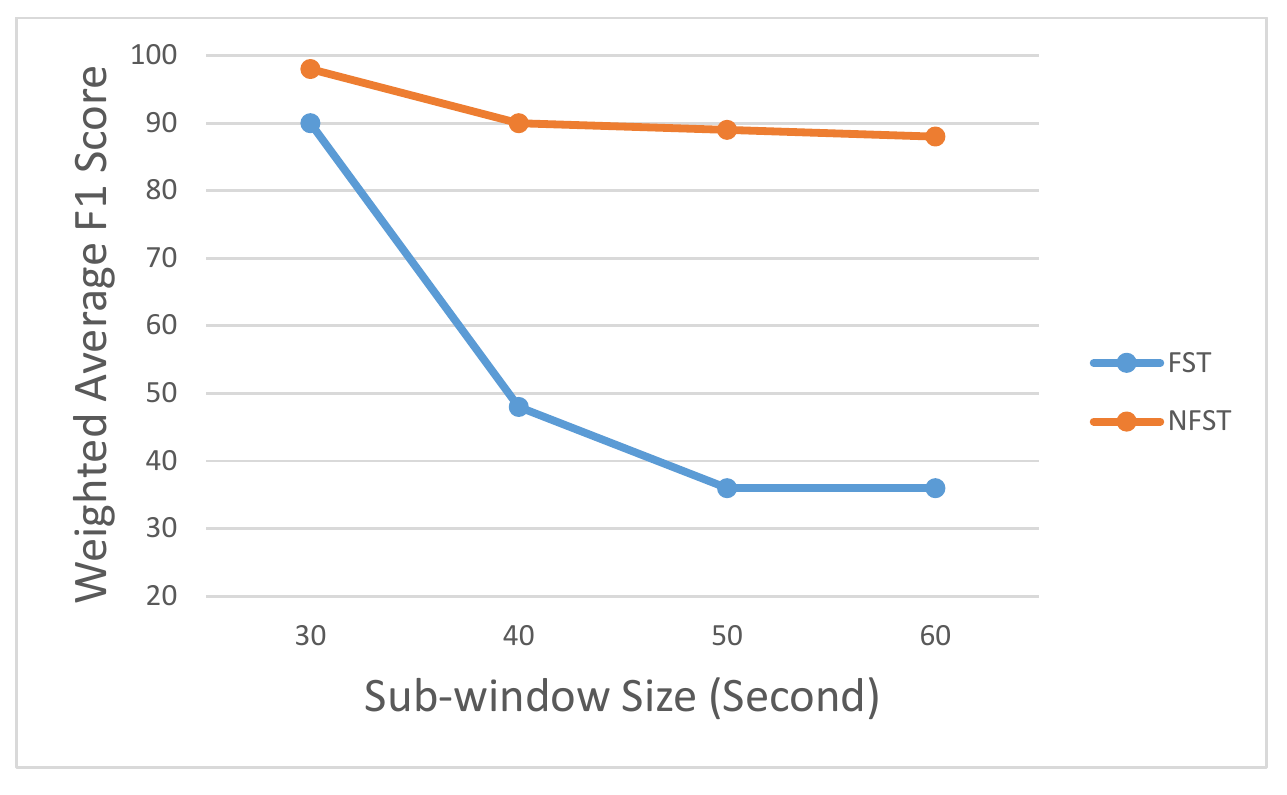}
\caption{Effect of the sub-window size on the RF model performance (F1-score)}
\label{rf_fst}
\end{figure}

We also analyzed the general effects of the windowing operation on the fatigue driving detection process. The performance results for diverse window sizes and each specific methodology are depicted in Figure~\ref{sub_window_effect}. 
The RF, KNN and SVC models show a trend of first increase and then decrease performance as the size of the window grows. Except the NB, the maximum performances are all obtained in the sub-window size of $30s$. 
The performance of LR increases monotonically throughout the change of diverse window sizes, while the performance of NB firstly increases (from $10s$ to $40s$) and then remains unchanged (from $40s$ to $60$). Both of the performance on LR and NB are not significantly influenced by the change of the sub-window size, and therefore, we designed a “cut-off” window size at $30s$ for feature extraction. Based on the experiment result, we observed that the RF model stands out among all evaluated models, which provides the highest performance, with an F1-score of $96\%$. However, increasing the window size to more than $30s$ drags down the detection performance more rapidly comparing with the other models. The reason is mainly due to the imbalanced samples between NFST and FST. Imbalanced training samples produce an important deterioration of the classification accuracy, in particular with the performances belonging to the less represented classes~\cite{barandela2004imbalanced}. In other words, the change of the sub-window size has a more significant impact (sensitivity) on the detection of FST (small class) than NFST (large class). For the models excluding the RF, they have a high accuracy on NFST but low accuracy on FST. Therefore, even changing the sub-window size is sensitive to the performance of FST, but because there is not much space to drop the accuracy of FST, the weighted average F1 score does not present a steeply decrease. Conversely, as the tree-based machine learning models work by learning a hierarchy of ``if/else'' questions and this can force both classes to be addressed, tree-based models usually perform well on imbalanced data. RF, as a tree-based model, has a high accuracy on both FST and NFST, and the change of sub-window size can easily drop the accuracy of FST a lot (from 90\% to 36\%), and therefore decrease the weighted average F1 score significantly (Figure~\ref{rf_fst}).

\subsubsection{Feature Analysis}
We are also interested in what are the important features for the fatigue driving detection and how these features impact the detection performance.

\begin{figure*}[htp]
\centering
\includegraphics[width=\linewidth]{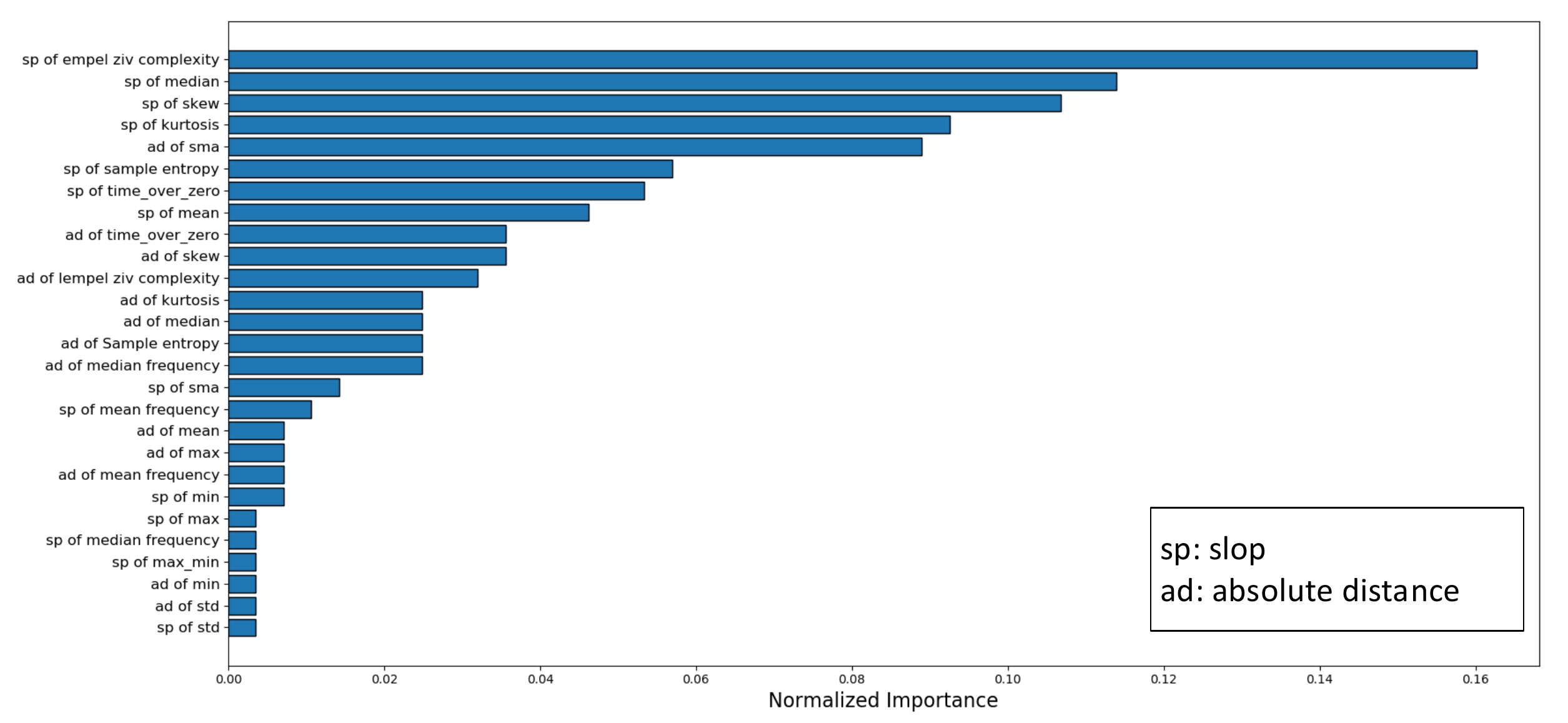}
\caption{The importance of the features generated in our system}
\label{feature_importance}
\end{figure*}

\vspace{2mm}
\noindent{\bf\em Feature Importance.} In this study, since we used the XGBoost to select the valid sEMG samples and Random Forest to detect the fatigue state transition (FST), in order to avoid the bias, we used another tree based model ``lightGBM''  to determine the feature importance. The 
importance of the features we generated is illustrated in Figure~\ref{feature_importance}, by which we can obtain the highly important features and lesser important features for FST detection.

\begin{figure}
    \centering
    \begin{subfigure}{0.5\textwidth}
    \centering
	\includegraphics[width=0.7\linewidth]{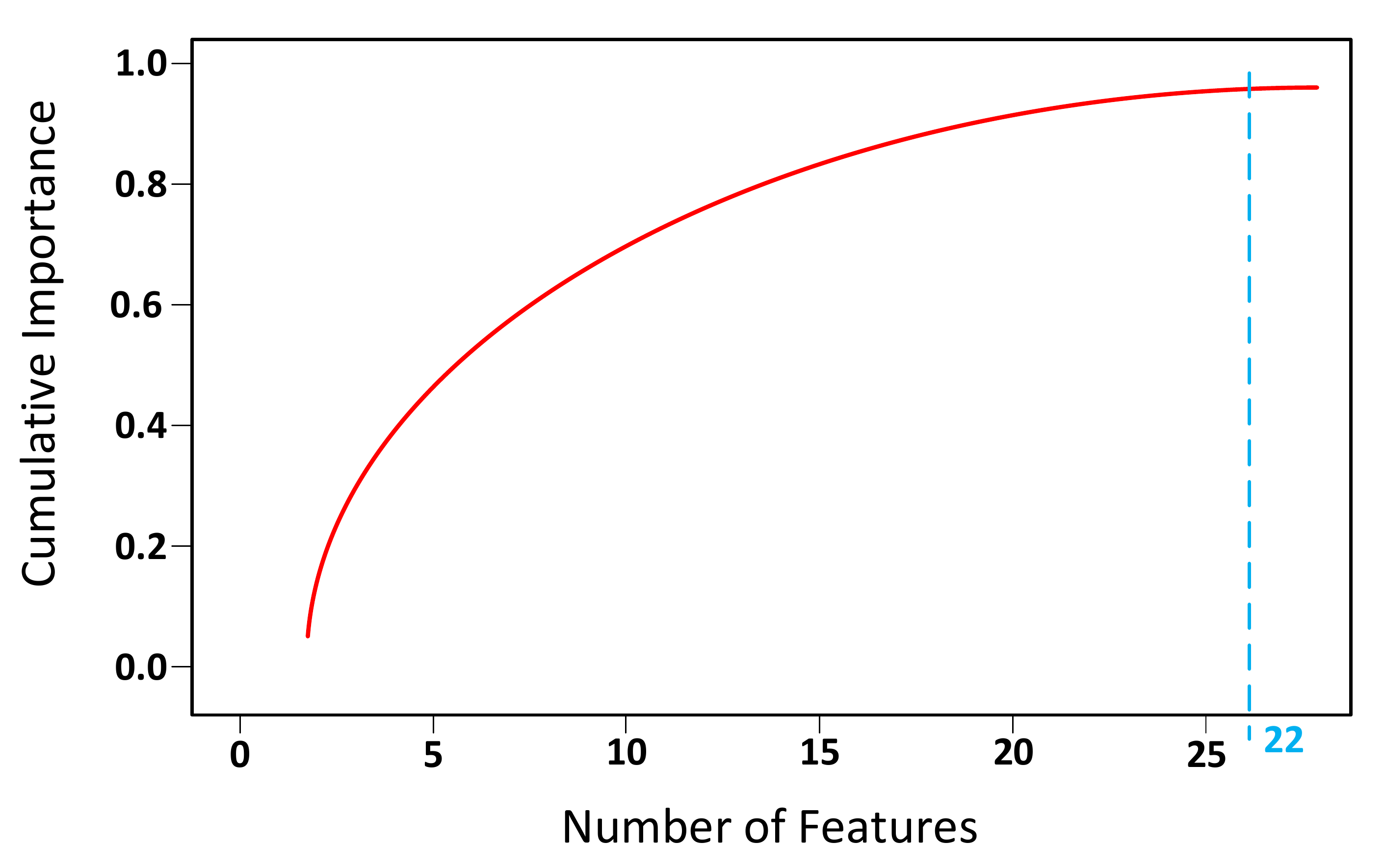}
\caption{Cumulative Feature Importance}
\label{cumulative_feature_importance}
    \end{subfigure}
    \begin{subfigure}{0.5\textwidth}
    \centering
	\includegraphics[width=0.8\linewidth]{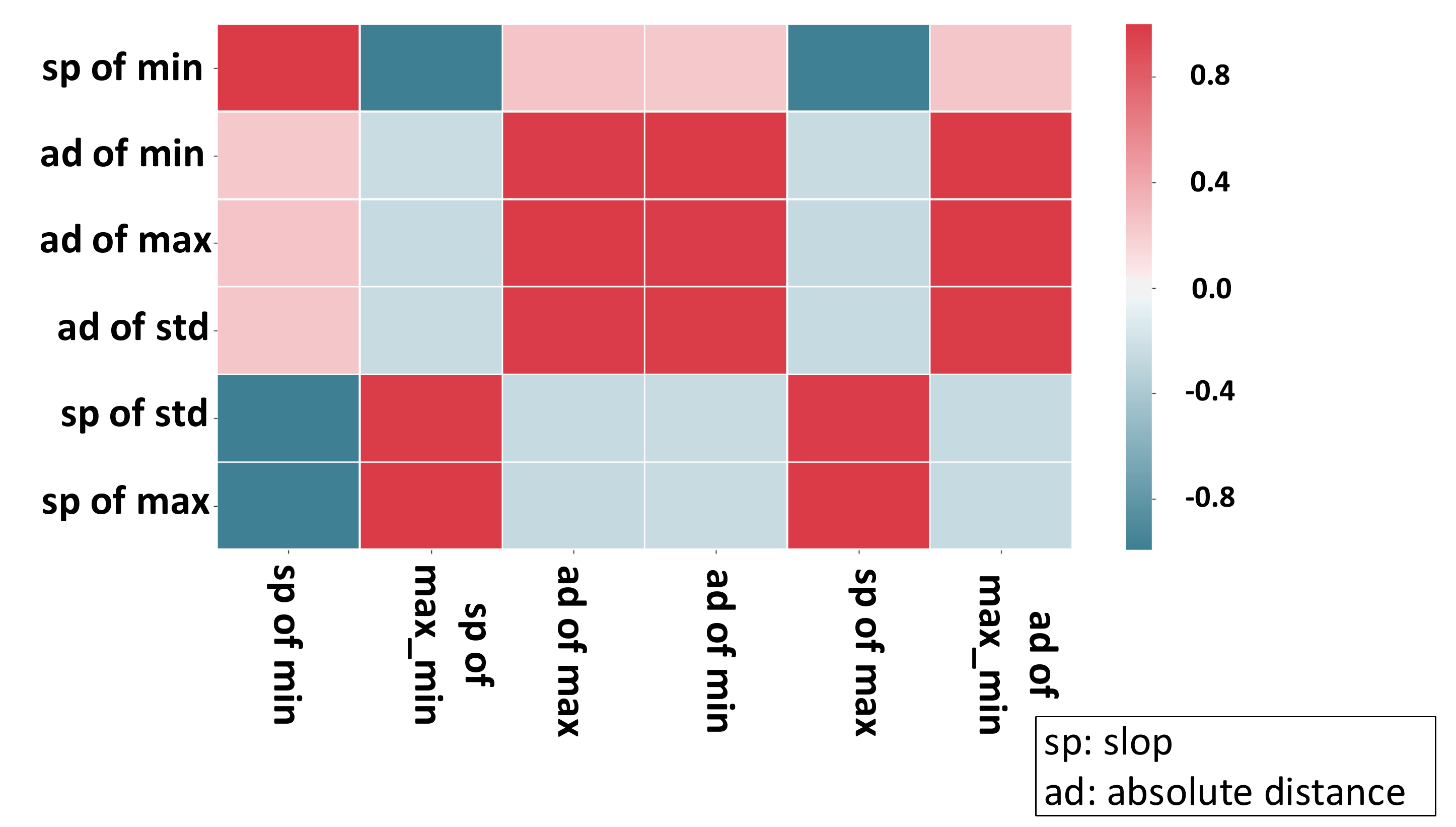}
\caption{Feature Correlations}
\label{correlation}
    \end{subfigure}
    \caption{Feature analysis}
    \label{fig:my_label}
\end{figure}

Slop of Empel Ziv Complexity, the slop of median, slop of Skew, slop of kurtosis and absolute distance of sma are the top 5 features which have significant contributions to detect FST, while slop of max, slop of median frequency, slop of max\_min, absolute distance of min, absolute distance of std and slop of std are less helpful to the FST detection. In Figure~\ref{cumulative_feature_importance}, the cumulative importance versus the number of features is illustrated. 

The vertical line is drawn at threshold of the cumulative importance, in this case is 90\%, which indicates that 23 features contribute 90\% effort for FST detection. This result gives a complement explanation to the fact that low important features do not contribute a lot in the fatigue state detection.



\vspace{2mm}
\noindent{\bf\em Feature Correlation.} In order to understand the reason why these features are low important to the FST, 
we conducted 
the correlation analysis using a heatmap. We find most of the lesser important features (excluding slop of median frequency) are collinear features which are highly correlated with one another, as shown in Figure~\ref{correlation}. In machine learning, this lead to decreased generalization performance due to high variance and less model interpretability. In a tree-based model, such as``lightGBM'', the collinear features are seldom or not used to split any nodes, and therefore these features are categorised as lesser important features eventually.

In order to further investigate the effect of the lesser important features, we removed 
those 
lesser important features and used the rest features to evaluate the fatigue state detection performance again. This time, only RF model was used due to its insensitivity on imbalanced samples. The weighted average F1 score and confusion matrix after removing the lesser important features under sub-window size of 30s are illustrated in Table~\ref{F1_remove} and Table~\ref{Confusion Matrix_remove}. The benefits of lesser important feature removal are reflected by an increase of F1 score from 96\% to 98\%, and a decrease on false positive (from $1$ to $0$) and false negative (from $6$ to $3$).

\begin{table*}[htp]
\caption{F1 score of RF after removing the row important features (sub-window size: 30s)}
\label{F1_remove}
\centering
\resizebox{\linewidth}{!}{%
\begin{tabular}{|m{1cm}|m{2cm}|m{15cm}|}
\hline
\textbf{Model} &\textbf{F1 score}  &\textbf{Parameters} \bigstrut\\ \hline
RF&\textbf{0.98}&bootstrap=True, class\_weight=None, criterion='entropy',
            max\_depth=20, max\_features='auto', max\_leaf\_nodes=None,
            min\_impurity\_decrease=0.0, min\_impurity\_split=None,
            min\_samples\_leaf=1, min\_samples\_split=2,
            min\_weigh\_fraction\_leaf=0.0, n\_estimators=15, n\_jobs=None,
            oob\_score=False, random\_state=None, verbose=0,
            warm\_start=False\bigstrut\\ \hline
\end{tabular}
}
\end{table*} 

\begin{table}[htp]
\caption{RF Confusion Matrix after removing the lesser important features (sub-window size: 30s)}
\label{Confusion Matrix_remove}
\centering
\resizebox{1\linewidth}{!}{%
\begin{tabular}{|m{2cm}|>{\centering\arraybackslash}p{3cm}|>{\centering\arraybackslash}p{2cm}|}
\hline
&Detected NFST&Detected FST\bigstrut\\ \hline
Actual NFST&142&0\bigstrut\\ \hline
Actual FST&3&35\bigstrut\\ \hline
\end{tabular}}
\end{table} 

\subsection{E2 Result}
\begin{figure}[t]
\centering
\includegraphics[width=3.5 in]{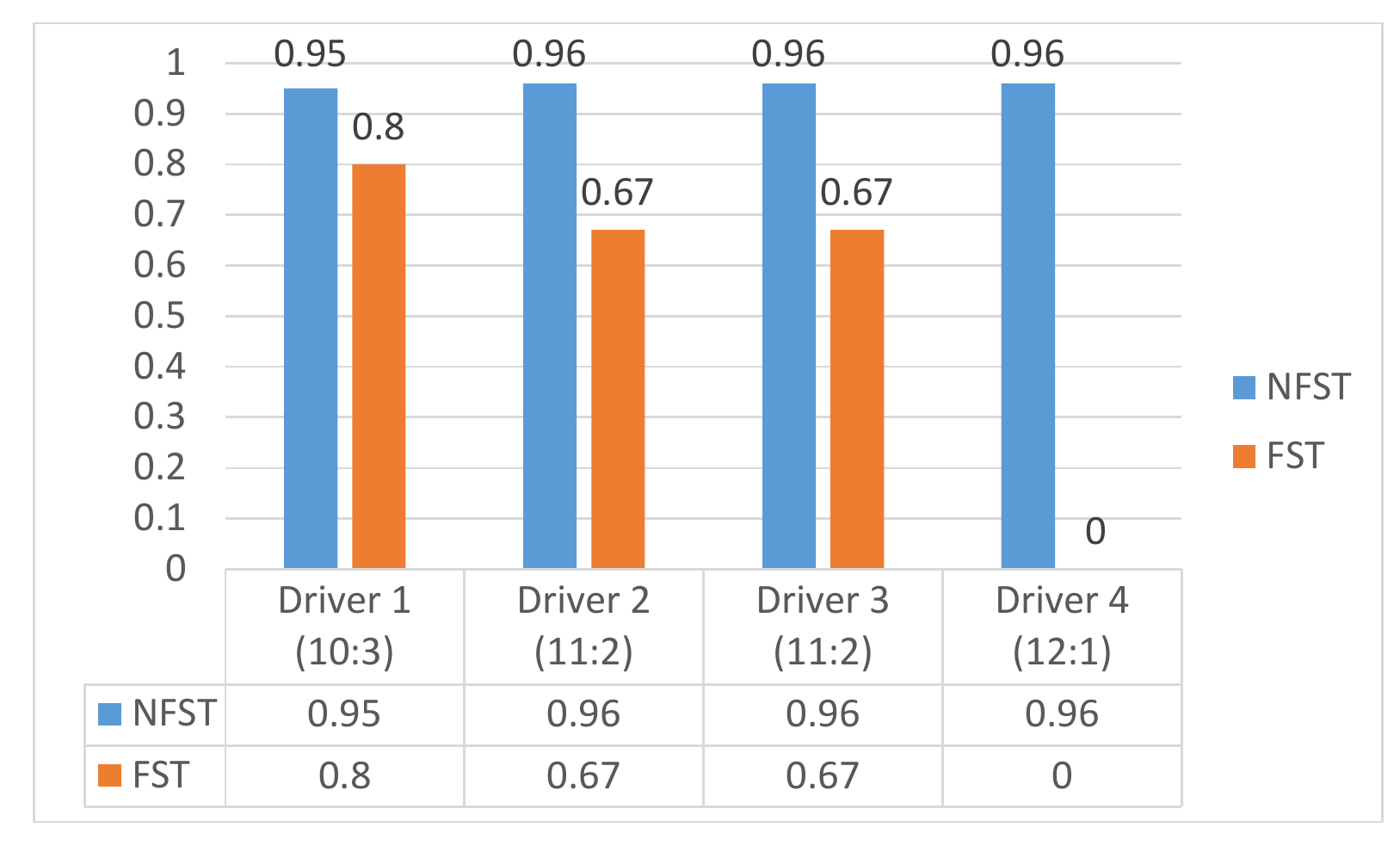}
\caption{NFST and FST detection results of RF using the feature set which is removed the lesser important features for E2}
\label{NFST_FST_RESULT}
\end{figure}

The data collected from real driving 
were 
used to test against the trained classification model in {\em E1}. The weighted average F1 scores of the four drivers are given in Table~\ref{F1 score for E2} and Figure~\ref{NFST_FST_RESULT}. From Table~\ref{F1 score for E2}, we can see, except the third driver's performance, the weighted average F1 scores generated from the feature set which is removed the lesser important features (fs1) are superior to the weighted average F1 scores generated from the whole feature set (fs2). More specifically, using fs1, the weighted average F1 scores for the first to third driver are all above 90\%, 
achieving 92\%, 91\% and 91\%, respectively. For the third driver, although the weighted average F1 score of fs1 (91\%) is lower than the weighted average F1 score of fs2 (100\%), the result is still acceptable. Moreover, Figure~\ref{NFST_FST_RESULT} illustrates the NFST and FST detection results of RF using the feature set which is removed the lesser important features, where driver 1 has 10 NFST ground truth and 3 FST ground truth, driver 2 has 11 NFST ground truth and 2 FST ground truth, driver 3 has 11 NFST ground truth and 2 FST ground truth, and driver 4 has 12 NFST ground truth and 1 FST ground truth, respectively. From Figure~\ref{NFST_FST_RESULT}, we can see the F1 socre regarding the detection of NFSTs and FSTs are $95\%$ and $80\%$ for the first driver, $96\%$ and $67\%$ for the second driver, $96\%$ and $67\%$ for the third driver, and $92\%$ and $0\%$ for the fourth driver, respectively. The result indicates that, except the fourth driver, most of the NFST and FST points can be detected using our solution. 

The weighted average F1 scores generated from the both feature sets for the fourth driver are worse than those for 
the other three drivers. The reason is mainly due to the additional/unusual driving behaviors of the fourth driver. During the experiment, we noted the first three drivers drove safely and turned smoothly. However, the fourth driver has been a professional Uber driver for more than $2$ years. He was used to driving for long hours and did not get tired in the first $100$ minutes of driving. This observation also reflected on the camera-based fatigue detection method, by which there was no highly confident result indicating he was under a fatigue state. During his driving, he only reported ``he felt a little bit fatigue'' at the end of the experiment. We labelled his report as FST, but the sEMG samples we collected at that FDP may not represent a real fatigue driving state corresponding to the FST. Furthermore, the fourth driver also had some unconscious habits to rub fingers with the sEMG sensors deployed on the driving wheel and sometimes held the bottom part of the steering wheel (the sEMG sensors were deployed on the upper part). All of these unusual driving behaviors of the fourth driver caused noises that led to the fairly low F1 score in Figure~\ref{NFST_FST_RESULT}. Overall, this real-world driving experiment shows encouraging results and useful insights to adopt our approach for real-life usage. 

\begin{table}[!htb]
\centering
    \caption{Overall F1 score of RF for E2}
    \label{F1 score for E2}
     \resizebox{1\linewidth}{!}{%
    \begin{tabular}
    {|m{2cm}|>{\centering\arraybackslash}p{3cm}|>{\centering\arraybackslash}p{2cm}|}
    \hline
     & F1 Score (feature set with lesser important features removed)  & F1 Score (whole feature set)   \\ \hline
    Driver 1  & 0.92 & 0.78  \\ \hline
    Driver 2  & 0.91 & 0.85   \\ \hline
    Driver 3  & 0.91 & 1.00  \\ \hline
    Driver 4  & 0.78 & 0.69   \\ \hline
   
\end{tabular}
}
\end{table}

\subsection{Discussion and Insights}
\label{sec:discussion}

\vspace{2mm}
\noindent{\bf\em Result Validity.}  
As per our study, the ``wave" 
forms 
from 
sEMG signals are observed whenever participants feel fatigue and intuitively change hand and body postures. The rise and drop of the corresponding sEMG feature values thus are crucial for us to detect FST. Yet, such forms may also be caused by hand movement, diverse noises or other confounding activities similar to the FST-correlated one. However, from our extensive studies in the two experiments, our approach is able to filter out the majority of these false ``wave" forms. Since there are much more NFST samples than FST samples in our experiments, we decided to evaluate both NFST and FST instead of just FST and introduce weighted average F1 to show more balanced evaluation results.

Our approach primarily uses ``wave-form'' to identify FST. However, it is still challenging to identify the correlation among FSTs. In our experimental settings, participants self-reported their fatigue level while driving. But in real life, drivers may take a break in the middle of driving or some energy drink in between, so that the FST detected afterwards might not be necessarily more severe than those detected before.

In E2, 
when drivers feel very tired after driving long hours, they may be too exhausted or even fall asleep without any reactive hand and body postures (no rise part of the ``wave" form). In such cases, our approach may not be able to detect FST. However, before drivers feel such a level of fatigue, they are very likely to have already exhibited multiple ``wave" forms for FST, which should have been successfully captured by our system. Thus, our system could effectively provide feedback in the real-life scenario and prevent drivers from entering into such an exhaustion status.

\vspace{2mm}
\noindent{\bf\em Insights.}
The use of two sEMG sensors deployed on the steering wheel may not be sufficient for all drivers. For example, the fourth driver in {\bf E2} has a personal habit of using bottom part of the steering wheel occasionally that impacted sensor data collection, hence, four sEMG sensors may need to be installed quarterly on the wheel. Also, plenty of noises 
are 
caused by finger rubbing of the sEMG sensors and the big movement of the wire that connects the data collection board and the sensors. Therefore, it is desirable to design a better hardware for noise-resilient sEMG sensors with a wireless connection to the board. A graphene-based sEMG sensor design solution is a potential research direction due to its mechanical flexibility and ultra-thinness 
high signal-to-noise ratio (SNR), and efficient signal transmission 
with the high electrical conductivity~\cite{huang2019graphene}.

Noises present in the obtained signals often compromised the accuracy of our proposed solution. It is challenging to remove such background noises entirely. This actually points out a promising research direction to use complementary sensors. That is, sEMG sensors can be leveraged with night-vision cameras and other bio-signals sensors, such as ECG devices, to minimize the noise impacts. In this direction, multimodal Deep Learning can be investigated further due to its ability to deal with the differences among complementary yet heterogeneous sensors with varying sampling rates, data types, and data format (discrete and continuous data)~\cite{radu2018multimodal}. 


%% file: conclusion.tex
\section{Conclusion}
\label{sec:conclusion}
In this paper, we present a novel approach to detect driving fatigue by deploying sEMG sensors on steering wheels. We design and build the sEMG sensors and data collection board, develop an innovative solution VeSEM to extract valid sEMG sensors from underlying diverse noises, and use multi-layer features to identify driver fatigue state transitions. Moreover, using comprehensive experiments involving 
17 participants on the simulated driving platform and on-road tests, we confirm and verify that our approach is able to detect driver fatigue state transition with an acceptable accuracy. Based on our experimental results, we raise a few important insights to push forward the state-of-the-art in detecting 
driving fatigue.

Our future work will focus on 
further 
improving the accuracy of our approach in the real-life scenarios. We are currently investigating the use of multi-view learning~\cite{xu2013survey} to leverage our sEMG sensors with  complimentary sensors (e.g., night vision camera and ECG sensors).